\documentclass[pdflatex,sn-mathphys-num]{sn-jnl}

\usepackage{graphicx}%
\usepackage{multirow}%
\usepackage{amsmath,amssymb,amsfonts}%
\usepackage{amsthm}%
\usepackage{mathrsfs}%
\usepackage[title]{appendix}%
\usepackage{xcolor}%
\usepackage{textcomp}%
\usepackage{manyfoot}%
\usepackage{booktabs}%
\usepackage{float}
\usepackage{algorithm}%
\usepackage{algorithmicx}%
\usepackage{algpseudocode}%
\usepackage{listings}%
\usepackage[T1]{fontenc}
\usepackage{tipa}
\usepackage{bm}
\usepackage{gensymb}
\usepackage{color, soul}
\usepackage[normalem]{ulem}

\usepackage[dvipsnames]{xcolor}

\begin{document}

\title[Article Title]{Non-Hermitian trapping of Dirac exciton-polariton condensates in a perovskite metasurface}

\author[1,2]{\fnm{Mikhail} \sur{Masharin}}\email{mikhail.masharin@epfl.ch}
\equalcont{These authors contributed equally to this work.}

\author[2]{\fnm{Igor} \sur{Chestnov}}\email{igor.chestnov@metalab.ifmo.ru}
\equalcont{These authors contributed equally to this work.}

\author[2,3]{\fnm{Andrey} \sur{Bochin}}

\author[3,4]{\fnm{Pavel} \sur{Kozhevin}}

\author[2,3]{\fnm{Vanik} \sur{Shahnazaryan}}

\author[2]{\fnm{Alexey} \sur{Yulin}}

\author[5]{\fnm{Ivan} \sur{Iorsh}}

\author[6]{\fnm{Xuekai} \sur{Ma}}

\author[6]{\fnm{Stefan} \sur{Schumacher}}

\author[2,7]{\fnm{Sergey} \sur{Makarov}}

\author[8]{\fnm{Anton} \sur{Samusev}}

\author*[3]{\fnm{Anton} \sur{Nalitov}}\email{anton.nalitov@gmail.com}

\affil[1]{\orgdiv{Institute of Bioengineering}, \orgname{\'Ecole Polytechnique F\'ed\'erale de Lausanne (EPFL)}, \orgaddress{ \city{Lausanne}, \postcode{1015}, \country{Switzerland}}}

\affil[2]{\orgdiv{School of Physics and Engineering}, \orgname{ITMO University}, \orgaddress{ \city{St. Petersburg}, \postcode{191002}, \country{Russia}}}

\affil[3]{\orgdiv{Abrikosov Center for Theoretical Physics},
\orgaddress{
\city{Dolgoprudnyi}, \postcode{141701},
\country{Russia}}}

\affil[4]{\orgdiv{Department of Physics, St. Petersburg State University},
\orgaddress{
\street{University Embankment, 7/9},
\city{St. Petersburg}, \postcode{199034},
\country{Russia}}}

\affil[5]{\orgdiv{Department of Physics, Engineering Physics and Astronomy}, \orgname{Queen’s Universiy}, \orgaddress{\city{Kingston}, \country{Canada}}}

\affil[6]{\orgdiv{Department of Physics and Center for Optoelectronics and Photonics Paderborn (CeOPP)}, \orgname{Paderborn University}, \orgaddress{\street{Warburger Strasse}, \city{Paderborn}, \postcode{33098},  \country{Germany}}}

\affil[7]{\orgdiv{Qingdao Innovation and Development Center}, \orgname{Harbin Engineering University}, \orgaddress{ \city{Qingdao}, \postcode{266000}, \country{China}}}

\affil[8]{\orgdiv{Experimentelle Physik 2}, \orgname{Technische Universit\"at Dortmund}, \orgaddress{

\city{Dortmund}, \postcode{44227}, 

\country{Germany}}}

\abstract{
Massless Dirac particles avoid trapping due to their exceptional tunneling properties manifested in the so-called Klein paradox. 
This conclusion stems from the conservative treatment, but so far, it has not been extended to a non-Hermitian framework.
Recently, driven-dissipative bosonic condensation of Dirac exciton-polaritons was demonstrated in metasurface waveguides.
Here, we report an experimental observation of spatial binding and energy quantization of Dirac exciton-polaritons in a halide perovskite metasurface. 
A combination of spatially profiled nonresonant optical excitation and exciton-polariton interaction forms an effective non-Hermitian complex potential responsible for the observed effect.
In the case of tightly focused pump spots spanning from 9 to 17~$\mu$m, several bound states simultaneously achieve macroscopic occupation, constituting a multi-mode bosonic condensation of exciton-polaritons. 
Our theoretical analysis based on the driven-dissipative extension of the Dirac equation reveals that the non-Hermitian character of the effective trap allows for confinement even in the case of the gapless Dirac-like photonic dispersion, both above and below the energy of the dispersion crossing.
}


\maketitle
The perfect transparency of any potential barrier for ultra-relativistic massless Dirac particles is, perhaps, the most vivid illustration of the Klein paradox \cite{klein1929reflexion}, raising subtle questions about the existence of energy quantization and confinement of relativistic particles \cite{Giachetti2008}.
With a gapless spectrum, continuum states always allow the particle to tunnel out of the potential well, making complete trapping impossible \cite{GreinerBook}. 
Although its direct demonstration in high-energy physics remains challenging, Klein tunneling can be effectively reproduced in other physical systems with Dirac-like excitations \cite{Lee2019,Zhang2022,Elahi2024}.
In particular, a condensed matter counterpart of this effect is found in electron interband tunneling in graphene, which prevents localization and normal incidence reflection at p-n junctions \cite{Katsnelson2006,Cheianov2007}.
Although backscattering is allowed in the two-dimensional case for laterally moving particles \cite{Silvestrov2007}, localization in potential wells remains suppressed due to intrinsic particle-hole symmetry \cite{Beenakker2008,CastroNeto2009}, and only quasi-bound states with finite lifetime exist \cite{Gutierrez2016}.
As the effect is manifested at the single-particle level, it can be equally observed in bosonic systems such as photonic \cite{Ozawa2017,Zhang2022} and phononic \cite{Jiang2020} crystals exhibiting Dirac cones in dispersion.

A peculiar case of bosons is presented by exciton-polaritons, mixed light-matter quasi-particles emerging from the strong coupling of photons and elementary excitations in semiconductors \cite{kavokin2017microcavities}.
Patterned planar optical cavities hosting polaritons offer a highly tunable platform for dispersion engineering. 
Dirac cones can be synthesized in artificial polaritonic honeycomb \cite{Jacqmin2014}, Lieb \cite{Klembt2017,Whittaker2018,Scafirimuto2021}, and kagome \cite{Harder2021} lattices allowing simulation of relativistic phenomena, including Klein tunneling \cite{Solnyshkov2016} and zitterbewegung \cite{Sedov2018,Lovett2023}.
Alternatively, strong excitonic interactions can be utilized to optically tailor desired dispersion \cite{Alyatkin2021,schmutzler2015all}.
Exciting polaritons with a spatially profiled non-resonant laser beam allows guiding and trapping them with optically induced effective potentials stemming from incoherent excitonic reservoirs \cite{Askitopoulos2013,Schneider2017}.
However, in addition to reservoir-induced potentials stemming from excitonic repulsion, polaritons experience position-dependent gain owing to bosonic-stimulated scattering.
It renders the created potentials complex-valued, hereby unveiling the intrinsically driven-dissipative nature of polariton systems.

Recently, non-equilibrium polariton condensation manifested in macroscopic occupation of a single quantum state with enhanced coherence was achieved in planar optical waveguides with subwavelength surface patterning (metasurfaces), where excitons are strongly coupled to optical bound-in-continuum modes \cite{Ardizzone2022,Grudinina2023,Masharin2023May,Wu2024}.
In this geometry, a subwavelength grating folds waveguide dispersion inside the light cone, resulting in a Dirac-like energy dispersion crossing at zero in-plane momentum. 
The condensate state can thus be captured with a non-Hermitian extension of a two-mode Dirac Hamiltonian \cite{Gianfrate2024,Sigurdsson2024Dirac}.
In contrast to the Hermitian case, where gapped dispersion arises from a grating-assisted diffractive coupling, non-Hermitian systems feature dissipative interaction between counter-propagating waveguide modes, analogous to spinor components of a Dirac fermion.
This type of interaction results in a level attraction, the phenomenon responsible for the gap collapse, accompanied by the emergence of two exceptional points \cite{Lu2020,Masharin2023May}.

Spatial localization and energy quantization of Dirac polaritons were recently observed in the state confined within the diffractive energy gap \cite{Riminucci2023}, the result consistent with a textbook problem of a massive Dirac particle in a potential box \cite{Alberto1996,Unanyan2009,GreinerBook}. 
The negative effective mass of lower-branch polaritons, rendering the optically induced reservoir potential effectively attractive, was thus highlighted as a key ingredient of trapping \cite{Nigro2023,Riminucci2023,Gianfrate2024}.  
This treatment, however, overlooks rich non-Hermitian physics exhibited by polariton systems.

In this paper, we revise the problem of potential trapping of Dirac particles from a non-Hermitian perspective. 
We demonstrate that, in stark contrast to the conservative case, where potential trapping beyond the energy gap is suppressed by Klein tunneling \cite{GreinerBook}, Dirac particles in the driven-dissipative framework can be effectively trapped with complex-valued potentials even in the case of a gapless energy spectrum.
The trapping effect is revealed in the size quantization of the exciton-polariton emission spectrum from perovskite methylammonium (MA) lead bromide
perovskite (MAPbBr$_3$) metasurface observed under pulsed, non-resonant optical excitation focused onto a few-micron pumping spot.
Indeed, the selected halide perovskite owns high exciton binding energy ($\sim$35~meV~\cite{Soufiani2015Dec}) and sufficiently high refractive index ($n > 2$), making it a promising platform for studying polariton nonlinear effects \cite{su2021perovskite, Alyatkin2021,fieramosca2025fully}.
In contrast to the conservative case, the level structure expands beyond the energy gap with level spacing about ten meV, exhibiting a signature of conventional energy quantization.
The spectral quantization, specific patterns of polariton emission in real and reciprocal spaces, and the dependence of polariton emission on pumping fluence, are in excellent agreement with a non-Hermitian driven-dissipative extension of the Dirac equation \cite{Sigurdsson2024Dirac}.

\section*{Results and Discussion}\label{sec:intro}

\begin{figure}[h!]
    \centering
    \includegraphics[width=0.98\linewidth]{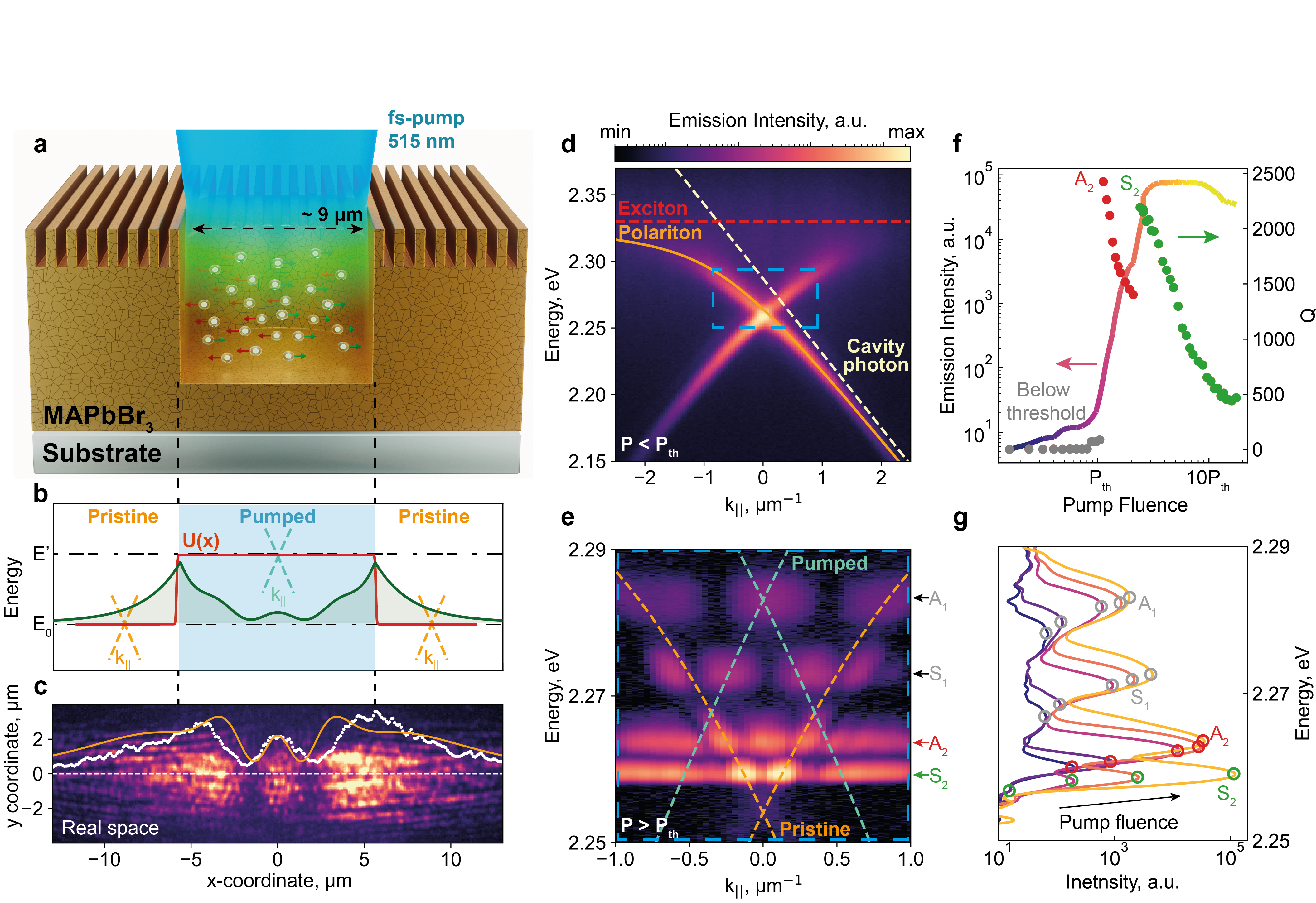}
    \caption{(a) A sketch of the locally pumped MAPbBr$_3$ metasurface and the effective optically induced non-Hermitian potential resulting in trapping and size-quantization of a nonequilibrium polariton condensate.
    (b) A model of the step-like non-Hermitian potential $U(x)$ (red line) and the condensate density profile (green solid line) for the highest growth rate state with the energy between the pristine $E_0$  (yellow) and blue-shifted $E^{\prime} = E_0 + {\rm Re}(U)$ (cyan) Dirac points of free polariton dispersion shown with dashed crosses.
    (c) Real-space image of the quantized polariton condensation emission. White dots represent the intensity profile along the $x$-axis, while the yellow line corresponds to the results of the full theoretical model.
    (d) Exciton-polariton energy dispersion (yellow solid line) emerging from strong coupling of exciton (red dashed line) and photonic waveguide (beige dashed line) modes in the measured angle-resolved emission spectrum under a femtosecond pump below the threshold pump fluence $P < P_{\rm th}$. The blue dashed rectangle illustrates the axes limits in panel (e).
    (e) Angle-resolved emission spectrum of a quantized polariton condensate above the pump threshold. The cyan and yellow lines schematically show polariton dispersion within and outside the pump region, respectively.
    (f) The total emission intensity as a function of pump fluence (solid colorized line, left vertical axis). The $Q$-factor, as a ratio of central energy $E_{i}$ and lasing linewidth $\Delta E$, of the dominant emitted mode is shown with circles (right vertical axis). The gray circles describe a broadband emission below threshold. 
    (g) The emission spectra integrated over the whole momentum $k_{\parallel}$ range for various pump fluences, demonstrating evolution of the quantized states. The colors correspond to the pump fluences shown in panel (f).
    }
    \label{fig1_v1}
\end{figure}

\subsection*{Experimental results}

To study spatial quantization of Dirac polaritons, we fabricate a one-dimensional rectangular-shaped metasurface composed of methylammonium lead bromide perovskite MAPbBr$_3$ as sketched in Fig.~\ref{fig1_v1}a. The details on metasurface fabrication and characterization are given in Methods.
A 90-nanometer-thin perovskite layer on a SiO$_2$ substrate supports waveguiding modes, while a sub-wavelength metasurface with 320~nm period and 20~nm thickness modulation folds their dispersion below the vacuum light cone (see Methods and Supplementary Figure~S1 for details). 

Strong confinement of optical modes within the active medium facilitates their hybridization with robust excitons \cite{Soufiani2015Dec}, which remain stable up to room temperature. 
The exciton-to-photon coupling strength of $38$~meV, extracted from the fitting of the angle-resolved photoluminescence (PL) spectrum with a coupled oscillators model \cite{Hopfield1958Dec}, confirms the formation of exciton-polariton metasurface modes with almost linear dispersion near the $\Gamma$-point (see Supplementary Section S1 and Supplementary Figure~S2 for details). 
There is no visible energy splitting between these counter-propagating modes, as in our case, where the grating fill factor is close to 0.5, the grating-assisted diffractive mode coupling is expected to be weak and can be hindered by mode broadening \cite{Masharin2023,ferrier2022unveiling}.
Despite this, the calculations show a strong dependence of the mode linewidths on the in-plane wavevector due to their coupling with the radiative continuum \cite{Ardizzone2022,Lu2020,Sigurdsson2024Dirac}.
The detailed analysis of the observed mode interaction is provided in Supplementary Section S1.

Under a mean-field treatment, such a momentum-resolved spectral behavior can thus be described with an effective non-Hermitian Hamiltonian expressed in the basis of the counterpropagating polariton states:
\begin{equation}
\hat{H}_{\rm D} = \left[
\begin{matrix}
\Omega_{\rm p} + \hbar v_g k_\parallel & V \\
V & \Omega_{\rm p} - \hbar v_g  k_\parallel
\end{matrix}
\right], \label{eq:hamilt}
\end{equation}
where $V$ describes the complex optical coupling between counterpropagating polaritons, $v_g$ is the group velocity and $\Omega_{\rm p}$ is the complex polariton energy at $\Gamma$-point ($k_{\parallel} = 0$). The extracted dispersion data (see SI, Section S1) fitted with Eq.~\eqref{eq:hamilt} confirm the dominance of the dissipative coupling, ${\rm{Im}}(V) \gg {\rm{Re}}(V)$, consistent with the collapse of the photonic gap at the $\Gamma$-point. 

The virtual absence of the photonic energy gap at the $\Gamma$-point gives the intuition that neither spatial confinement nor energy quantization are expected in our system. Indeed, a Dirac particle can only be trapped in a potential well within the forbidden energy range, that is, in the gap \cite{GreinerBook}. Otherwise, it can always escape the well via Klein tunneling (see Supplementary Section~S4.1 for details). 
To test this hypothesis, we pump the sample non-resonantly with femtosecond pulses focused on a few-$\mu m$ spot and measure the fluence-dependent angle-resolved emission spectra.
The interband excitation creates an incoherent background of charge carriers and high-energy excitons within the pump spot \cite{Masharin2024Jan}, locally shifting the polariton dispersion up in energy, as shown schematically in Fig.~\ref{fig1_v1}b.
At pump fluences above the condensation threshold, polariton trapping occurs, manifested by a speckled emission pattern in real space with a structured distribution of condensed polaritons (Fig.~\ref{fig1_v1}c).

At the same time, in the angle-resolved emission spectra approaching the pump threshold, we observe a transition from the smoothly distributed polaritons over the branch into a ladder of well-resolved energy high-intensity peaks, indicating a strong size-quantization effect (Figs.~\ref{fig1_v1}d-e). 
Remarkably, the energy quantization becomes visible below the polariton lasing threshold (see Supplementary Section S2 for the detailed discussion).
In addition, in analogy to the particle-in-a-box problem, the detected $k_{\parallel}$-space emission features alternating symmetry of energy levels.
This symmetry can thus be utilized for classification of the emerged states: the states with the emission peak at $k_{\parallel}=0$ are identified as antisymmetric while those with a minimum at zero momentum are symmetric, see the Methods and SI Section~S5. 
In the momentum-integrated spectrum (Fig.~\ref{fig1_v1}g), they correspond to the peaks $A_1$, $S_1$, $A_2$, and $S_2$ arranged from top to bottom.
Notably, at large pump spots (>~30~$\mu m$) the size quantization effect vanishes, and single-mode polariton condensation occurs, as was demonstrated in Refs.  ~\cite{Masharin2024Jan, Masharin2023May}.

The pump fluence threshold of 1.27~mJ/cm$^2$ is evidenced by a superlinear increase in emission intensity, accompanied by a pronounced narrowing of the spectral linewidth characterized by the incease of the emission peak's $Q$-factor, which stands for the ratio of central emission energy and its linewidth: $Q = E_{i}/\Delta E$, as shown in Fig.~\ref{fig1_v1}f.
At pump fluences between $P_{\rm th}$ and $1.5P_{\rm th}$, the $A_2$ mode dominates. Above $1.5P_{\rm th}$, the $S_2$ mode emerges and becomes dominant in both intensity and $Q$-factor.
This behavior is also reflected in the emission intensity as a function of pump fluence dependence (L-L curve), which exhibits a kink around $\sim 1.5P_{\rm th}$ (Fig.~\ref{fig1_v1}f).

The maximal achieved value of the $Q$-factor of the dominant mode reaches 2450. However, as the pump fluence increases, the modes broaden due to the rapid temporal evolution of polariton and reservoir densities under pulsed excitation, as well as due to sample heating.
The instantaneous spectral position of the quantized states continuously shifts in response to the time-varying potential and additionally undergoes a time-dependent blueshift, resulting in spectral broadening \cite{Masharin2024Jan}.
This effect is also observed in the momentum-integrated spectra (Fig.~\ref{fig1_v1}g), as well as in the pump-fluence dependent blueshift of the quantized modes, primarily caused by polariton–polariton interactions.

\begin{figure}
    \centering
    \includegraphics[width=1.0\linewidth]{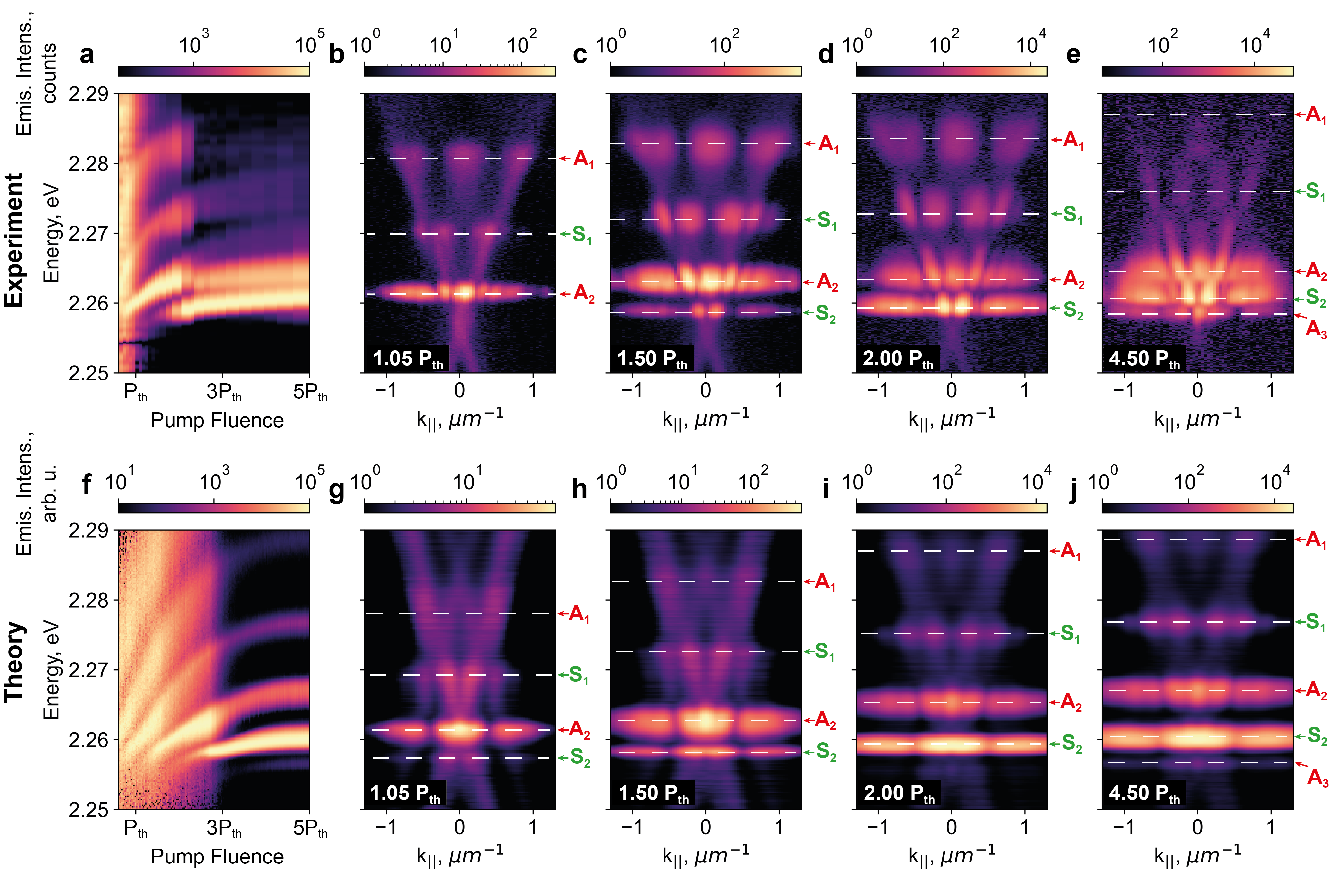}
    \caption{Integrated over the angles, measured (a) and numerically modeled (f) emission spectra as a function of pump fluence. Measured angle-resolved emission spectra at pumps of 1.05, 1.5, 2.0, and 4.5 of $P_{\rm th}$ (corresponding to 1.32, 1.94, 2.64, and 5.82 ${\rm mJ}/{\rm cm}^2$ respectively) (b-e) and numerically modeled angle-resolved emission spectra at the corresponding pump strengths (g-j). 
    }
    \label{fig2_new}
\end{figure}

To investigate the evolution of the energy structure, we measured momentum-integrated emission spectra as a function of pump fluence (Fig.~\ref{fig2_new}a), as well as angle-resolved emission spectra at pump fluences of $1.05P_{\rm th}$, $1.5 P_{\rm th}$, $2 P_{\rm th}$, and $4.5 P_{\rm th}$ (Figs.~\ref{fig2_new}b–e, respectively).
In the vicinity of the threshold fluence $P_{\rm th}$, the emission pattern is characterized by three quantized states $A_1$, $S_1$, and $A_2$ as shown in Fig.~\ref{fig2_new}b.
Among them, the lower-energy $A_2$ mode dominates in the emission intensity.
The collected spectra successfully capture a subsequent switch to the emergent symmetric $S_2$-mode near the Dirac Point at $~1.5$ times the threshold (Fig.~\ref{fig2_new}c) as well as a blushift and level broadening (Fig.~\ref{fig2_new}d), consistent with the behavior discussed above.
Moreover, the $A_3$ mode appears at the pump fluence around $4.5P_{\rm th}$ (Fig.~\ref{fig2_new}e). However, due to the sample photodegradation limitation, its subsequent evolution could not be investigated.

\subsection*{Discussion}

We attribute the surprisingly strong effect of energy quantization of Dirac-like excitations to the non-Hermitian nature of polaritons. 
In what follows, we develop a detailed description of the confinement of Dirac polaritons. 
To quantitatively reproduce the experimental data, we resort to dynamical mean-field simulations based on the driven-dissipative generalization of the Gross-Pitaevskii equation supplemented with the kinetic equation for the incoherent reservoir density \cite{Carusotto2013,Sigurdsson2024Dirac}.
Details of the calculations are reported in Methods and SI Section S3.

The trapping effect stems from the tightly focused optical pump responsible for an effective complex potential for Dirac polaritons:
\begin{equation} \label{eq:potential}
   U = (\alpha + i \beta )N.
\end{equation} 
Here, $\alpha$ describes the cumulative strength of interaction with incoherent excitons characterized by the density $N$ of excitons, subject to bosonic-stimulated relaxation to the polariton states with the rate $\beta$. 
The potential shape follows the pump beam profile, while its height is proportional to the pump strength.

The incoherent emission, which dominates the photoluminescence spectrum below threshold, is simulated by adding stochastic noise associated with spontaneous scattering from the reservoir. Above threshold, the noise term is responsible for seeding multimode emission. 
Under a continuous-wave excitation, the stimulated relaxation from the reservoir favors single-mode condensation as shown in Ref.~\cite{Riminucci2023}. 
However, under ultra-short excitation conditions, the mode competition does not have enough time to suppress multimode emission, as predicted by our model.

Our numerical simulations of the Dirac polariton condensation dynamics under femtosecond optical pumping (Eqs. \eqref{eq:DynModel} in Methods, see also Supplementary Section S3) show excellent agreement with experimental results. 
In particular, it accurately reproduces the $k_{\parallel}$-space emission patterns of the quantized levels, along with the pump-fluence-dependent evolution of their spectral position (Figs.~\ref{fig2_new}g–\ref{fig2_new}j).
The real-space distribution of a far-field condensate emission shown in Fig.~\ref{fig1_v1}c is also in good agreement with the real-space image of the emission directly acquired in the experiment.
In addition, our simulations predict the cascaded switching of the dominant lasing mode, confirmed by our measurements.

The good correspondence between the full theoretical model of Dirac polariton condensation and experiment  (Fig.~\ref{fig2_new}) regarding the energy structure and wave functions gives us confidence to explore further the problem of non-Hermitian confinement of Dirac particles.
To reveal the fundamental mechanism behind the observed energy quantization, we proceed by simplifying the theoretical framework.

We further study the formation of bound (localized) states by solving the eigenvalue problem governed by a non-Hermitian generalization of the 1D Dirac Hamiltonian \eqref{eq:hamilt} in the presence of a complex potential \eqref{eq:potential} (see Supplementary Section S4 for details):
\begin{equation}
\left[\hat{H}_{\rm D} + U(x)\right]\vec{\Psi}(x) = E\vec{\Psi}(x),
\label{eq:eigenValueProblem}
\end{equation}
where $\vec{\Psi} = \left(\Psi_+, \Psi_-\right)^\intercal$ is the pseudo-spinor condensate wavefunction in the basis of forward- and backward-propagating polariton states. 
For simplicity, we assume a box-shaped potential profile defined by the incoherent reservoir density $N(x) = N_0$ within the pumped region ($|x| < D$, where $2D$ is the pump spot size) and $N(x) = 0$ outside, as sketched in Fig.~\ref{fig1_v1}b. 

The eigenvalue equation \eqref{eq:eigenValueProblem} thus reduces to a non-Hermitian two-component extension of the quantum well problem.  Requiring continuity of both components of the wavefunction across the piecewise-defined regions, and specifically at the borders of the well, $|x|=D$, we determine the energy eigenvalues $E_n$ (see SI for the calculation details).
Due to the intrinsic non-Hermiticity of the problem, solutions with exponentially decreasing tails beyond the trap have complex energies and therefore can be classified as \textit{quasi-bound} states.
The real parts of $E_n$ define the spectral positions, while the imaginary parts indicate the temporal behavior of the modes: amplification at $\mathrm{Im}( E_n )>0$ and decay in the case of $\mathrm{Im}( E_n )<0$. The state with the largest imaginary energy will, therefore, dominate the emission spectrum. At the same time, all the other amplifying modes are expected to be macroscopically populated during the rapid condensate formation and are thus also visible in the emission spectrum.

\begin{figure}
    \centering
    \includegraphics[width=1.0\linewidth]{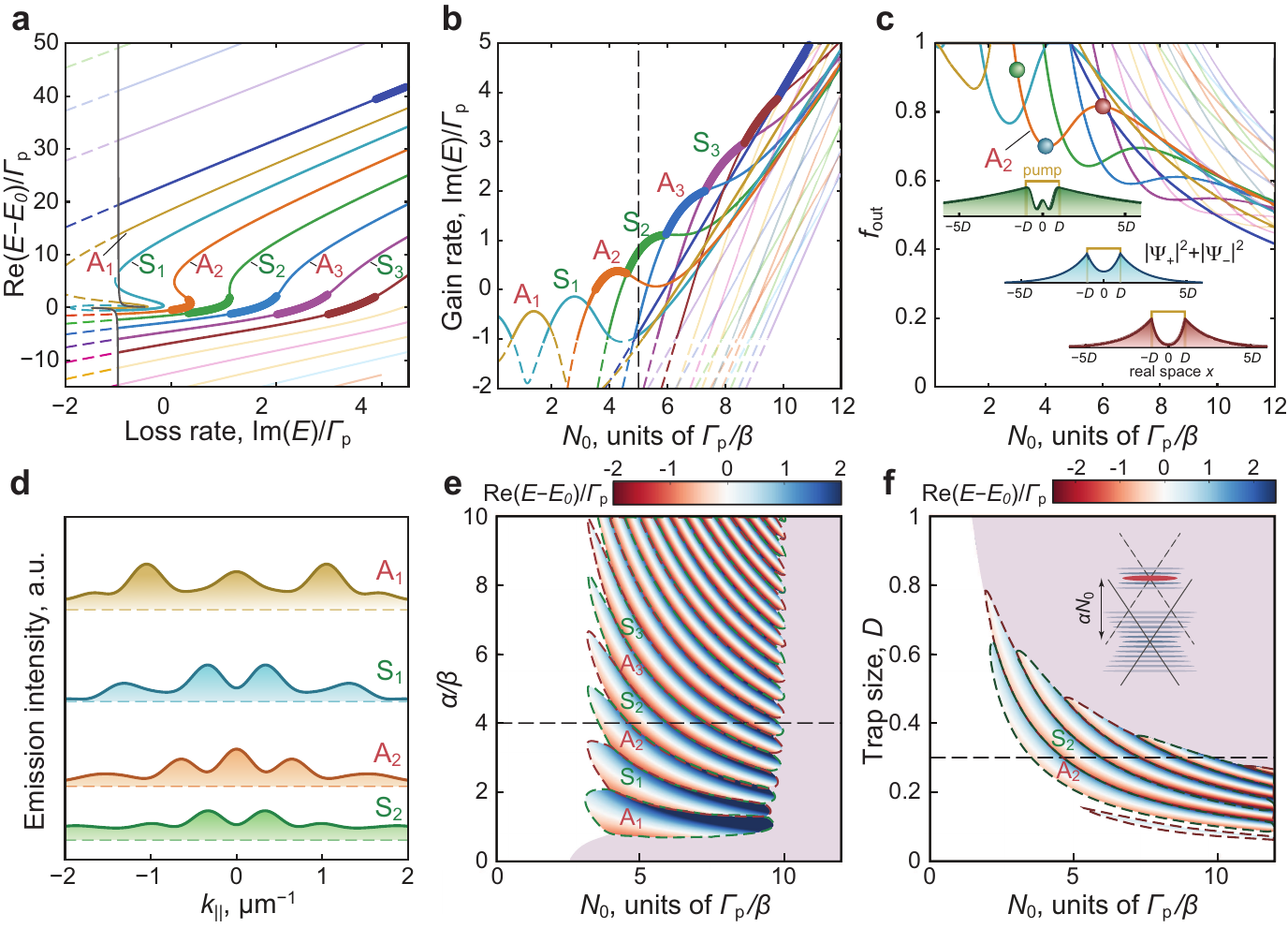}
    \caption{(a) The complex energy diagram of the eigenvalue problem \eqref{eq:eigenValueProblem}. The parameters are $\alpha/\beta = 4$, $D=0.3\hbar v_g/\Gamma_{\rm p}$, $V=(0.1+0.5i)\Gamma_{\rm p}$, where $\Gamma_{\rm p}$ is the polariton linewidth. The energy level is counted from the Dirac point position $E_0 \approx 2.26$~eV.
    (b) The imaginary part of the complex eigenenergies versus peak reservoir density $N_0$ measured in units of $\Gamma_{\rm p}/\beta$. 
    The dashed segments correspond to the solutions with exponentially divergent wavefunctions. The thick segments mark the state with the highest growth rate, which dominates in the emission spectrum. 
    (c) Localization efficiency criterion $f_{\rm out}$ describing the fraction of polaritons localized outside the trap. 
    The inset shows polariton densities of the state $A_2$ for the three values of potential height $N_0$ marked with color dots.
    (d) Reciprocal space cross-sections of quantized levels at $N_0\beta/\Gamma_{\rm p}=5$ (the gray dashed line in panel (a)); cf. with Fig.~\ref{fig1_v1}e. 
    (e) The phase diagram of the state with the highest growth rate, which dominates in the emission spectrum. The vertical axis shows the ratio between the repulsion $\alpha$ and gain $\beta$ parts of the complex trapping potential \eqref{eq:potential}. The color shows the real part of the dominating mode eigenenergy measured in units of $\Gamma_{\rm p}$.
    (d) The phase diagram on the $(D,N_0)$ parameter space.
    In (e) and (f), all the states are below threshold within the white region, while in the pale-red region, the condensation occurs near the blue-shifted Dirac point as sketched in the inset to panel (f). The dashed horizontal lines indicate the parameters corresponding to panels (a)-(c), relevant to our experimental conditions.}
    \label{fig3_new}
\end{figure}

Despite its simplicity, the eigenvalue problem \eqref{eq:eigenValueProblem} provides an intuitive description of all the observed phenomena, including size quantization and the pump-fluence-dependent evolution of the emission spectrum.
The bound-state solutions of Eq.~\eqref{eq:eigenValueProblem}, found using experimentally relevant parameters, form an energy ladder with real energy parts being approximately equidistant (see Supplementary Figure~S7 for details).
Remarkably, the bound states appear on both sides of the Dirac crossing point, i.e., from both positive and negative energy branches, as shown in the complex energy diagram Fig.~\ref{fig3_new}a.
Moreover, the trapping efficiency quantified by the fraction of polariton density localized outside the trap, $f_{\rm out}$, is independent of whether the state is located above or below the Dirac point (Fig.~\ref{fig3_new}c).
Therefore, the sign of the polariton effective mass is an insufficient factor for non-Hermitian localization, in contrast to a general intuition based on analogies with the conservative Schr{\"o}dinger equation \cite{Nigro2023,Riminucci2023}.

The bound states are formed with an effectively gapless dispersion. Therefore, these states differ from the 'gap-confined' solutions \cite{Riminucci2023}, bound because of their decay outside the trap in the energy range forbidden for free particle propagation. 
Such states provide a complete set of solutions in the conservative limit \cite{GreinerBook}. 
Non-Hermiticity generalizes this picture by expanding the eigenvalue problem into the complex energy domain, thereby relaxing the resonance condition imposed by the wavefunction continuity requirement. 
Localization then requires the positive imaginary part of the free-particle wave vector $q(E) = \sqrt{E^2 - V^2}/(\hbar v_g)$, ensuring that the wave function $\vec{\Psi} \propto \exp{\left(iq|x|\right)}$ decays outside the trap (see Supplementary Section S4 for details). 
Otherwise, the state is unbound -- the dashed segments of the lines in Figs.~\ref{fig3_new}a-c.
It is worth mentioning that in the system under study, non-Hermiticity stems from several different sources, including radiative losses, dissipative coupling, and spatially localized optical pumping.
Our analysis reveals that it is the local gain that should be recognized as the main factor of the observed non-Hermitian confinement, as discussed in Supplementary Section S4.

The reliability of the simplified trapping model is confirmed by a good correspondence between the predicted emission intensity patterns (Fig.~\ref{fig3_new}d) and the measured spectral distributions shown in   Fig.~\ref{fig1_v1}e.
In addition, the eigenvalue problem sheds light on the origin of the observed cascade of mode-dropping transitions. 
Figure~\ref{fig3_new}b shows a strongly nonmonotonous behavior of ${\rm{Im}}(E)$ responsible for the loss rate. This nonmonotonicity stems from the variation of the polariton density distribution as shown in Fig.~\ref{fig3_new}c, focusing on the $A_2$-state as an example.
At weak potentials, the state is weakly delocalized with a high fraction $f_{\rm out}$ of ballistic tails outside the trap. The increase in gain initially favors the density of the polaritons to develop mainly inside the trap, and $f_{\rm out}$ falls. However, the increasing role of exciton repulsion then pushes polaritons outside. The weight of radiative losses of polaritons that escape the trap is enhanced, resulting in a drop in the mode gain.  

At sufficiently strong repulsion $\alpha/\beta \gtrsim 1$, this behavior results in a cascade of successive transitions, in which the fastest growing mode remains in the vicinity of the Dirac point energy (see Fig.~\ref{fig3_new}a), in agreement with our experimental findings.
The universality of this mechanism is summarized in the energy diagram in Fig.~\ref{fig3_new}e. Note that beyond a critical reservoir density, the highest growth rate is expected to switch to the state located at the top of the trapping potential near the blue-shifted Dirac point energy $E_0+\alpha N_0$, see Fig.~\ref{fig3_new}f. This switching transition remained unobserved in our experiment because of the material photodegradation constraints. Remarkably, our experimental observations for different pump spot sizes confirm the predicted behavior (see Supplementary Figure~S5).

The cascade mode switching is expected to appear if the exciton repulsion dominates the stimulated scattering rate and under a sufficiently small pump spot (see Fig.~\ref{fig3_new}\sout{f}e). Our full model simulations, shown in Fig.~\ref{fig2_new}, together with the stationary problem analysis, both provide a good coincidence with experimental results at the $\alpha/\beta$-ratio of about 4.

In conclusion, we have demonstrated the non-Hermitian trapping and energy quantization of exciton-polaritons in a perovskite metasurface with a gapless Dirac energy spectrum, a phenomenon forbidden in the conservative regime by the Klein paradox.
Experimentally, we observed the formation of a multimode bosonic condensate under focused optical pumping, evidenced by a ladder of quantized energy states in the emission spectrum both above and below the condensation threshold. 
Alternating parity of reciprocal space emission patterns from the size-quantized states manifests the intrinsic symmetry of the non-Hermitian Dirac Hamiltonian and vividly illustrates the energy quantization effect.
Our theoretical framework, based on a driven-dissipative extension of the Dirac equation, reveals that the key mechanism for this confinement is the optically induced complex effective potential. 
This work reveals non-Hermiticity as the previously overlooked paradigm for confining and controlling ultrarelativistic quasiparticles with gapless energy dispersions, opening new avenues for engineering topological and quantum fluids of light.

\section*{Methods}\label{sec13}

\subsection*{Sample fabrication}

The sample is fabricated by spin-coating with subsequent nanoimprint lithography \cite{Makarov2017Apr, Masharin2022Nov}. The fabrication process comprises three primary stages: solution preparation, thin-film spin coating, and nanoimprint lithography. At the first stage, a perovskite solution was prepared in a dry nitrogen glovebox by dissolving 33.59~mg of MABr (Tokyo Chemical Industry, TCI) and 110.1 mg of PbBr$_2$ (TCI) into a 1 mL solvent mixture of DMF and DMSO at a 3:1 volume ratio. This solution was then stirred with a magnetic stirrer for 24 hours.

Next, the SiO$_2$ substrates were cleaned to ensure surface purity. This cleaning process involved sequential sonication in deionized water, acetone, and 2-propanol for 10 minutes each. After drying, the substrates were treated in an oxygen plasma cleaner for 10 minutes. The cleaned substrates were then returned to the dry nitrogen glovebox for film deposition.

The perovskite film deposition began with the deposition of 30 $\mu$L of the MAPbBr$_3$ solution onto each cleaned substrate. The spin-coating process was initiated, spinning the substrate for 50 seconds at 3000 rpm. At the 35-second mark, 300 $\mu$L of toluene was gently dispensed onto the rotating substrate to act as an antisolvent and promote rapid, uniform crystallization of the film.

Finally, the perovskite-coated substrate, still in its intermediate crystal phase before annealing, was transferred to a laboratory press for nanoimprint lithography. The nanoimprint mold featured a periodic grating structure with a period of 320 nm, a comb height of 20~nm, and a width-to-period ratio (fill-factor) of 0.48. The substrate, aligned with the mold, was subjected to approximately 200 MPa of pressure for 10 minutes. Following the mold's removal, the imprinted perovskite film underwent thermal annealing at 90${\degree}$C for 10 minutes in the dry nitrogen glovebox.

\subsection*{Experimental setup}

Pump-dependent emission spectra with angular resolution were obtained using a back-focal-plane (BFP) imaging setup that incorporated a slit spectrometer (Princeton Instruments SP2500) and a liquid-nitrogen-cooled CCD camera (PyLoN, Princeton Instruments). The photoluminescence (PL) and lasing were studied with non-resonant excitation from a femtosecond laser system (Pharos, Light Conversion) coupled to a second-harmonic generation crystal. The central wavelength of the laser emission was 515~nm with a pulse repetition rate of 100 kHz.

To ensure precise beam focusing and collection of emission signals, an infinity-corrected objective (Mitutoyo 100× VIS HR, NA = 0.90) was used. Samples were mounted in a holder within a sealed box, connected to a pump system to exclude photoinduced oxidation of the perovskite during the experiments. To control the pump spot size, the laser beam was focused by a 1000-mm-focus lens into the BFP of the objective, thereby limiting the angle of incidence and increasing the pump spot size.

A spatial filter in the intermediate image plane was implemented to collect the emission precisely from the pump spot center and to eliminate parasitic signals and reflections originating from the optical components, thereby enhancing signal fidelity.

\subsection*{Numerical simulations}
The multimode condensation phenomenon is described by the extended Gross-Pitaevskii equation:
\begin{subequations}\label{eq:DynModel}
\begin{align}
    \partial_{t} n_{\rm d} &= P_{\rm d}  - \gamma_{\rm r} n_{\rm d},\label{eq:nd}\\
    \partial_{t} n_{\rm b} &= P_{\rm b}  - (\gamma_{\rm r} + R \vec{\Psi}^\dagger \vec{\Psi} ) n_{\rm b}, \label{eq:nb}\\
    i\hbar \partial_{t} \vec{\Psi} &= \left( \hat{H}_{\rm D}  + U(n_{\rm b}, n_{\rm d}) + g\vec{\Psi}^\dagger \vec{\Psi} \right) \vec{\Psi} + \sqrt{D} \vec{W}(x,t), \label{eq:GPE}
\end{align}    
\end{subequations}
where $n_{\rm b}$ and $n_{\rm d}$ describe optically active (bright) and inactive (dark) exciton states, respectively, created with rates $P_{\rm b,d}(x,t)$ which follow a Gaussian real-space shape of the pump beam and a 220~fs FHWM temporal profile. The reservoirs decay with the rate $\gamma_{\rm r}$, but it is only bright excitons that are scattered to the condensate at the rate $R$. 

The linear Dirac Hamiltonian $\hat{H}_{\rm D}$ is given by Eq.~\eqref{eq:hamilt}, while the non-Hermitian trapping potential $U$ follows spatial distribution of the incoherent exciton fraction, $ U(n_{\rm b}, n_{\rm d}) = g_{\rm r} \left(n_{\rm b} + n_{\rm d}\right) + i/2 R n_{\rm b}$. Here $g_{\rm r}$ is the exciton-exciton scattering constant assumed to be identical for the bright and dark excitons. Polariton self-interaction parameter is defined as $g = g_{\rm r}X_{\mathrm{p}}^2$, where $X_{\mathrm{p}}^2$ is the exciton fraction (Hopfield coefficient). Stochastic noise is modeled as a Wiener process $\vec{W} = (W_+,W_-)^\intercal$ with spectral density $\langle W_i(x,t) W_j(x^\prime,t^\prime) \rangle =(Rn_{\rm b} + 2\Gamma_{\rm p})/(2dx) \delta(t-t^\prime) \delta_{i,j}\delta_{x,x^\prime}$, where $(i,j) = (+,-)$ and $dx$ is the lattice constant of the discretized mesh.

The following values of the system parameters were used: the photonic mode coupling strength $V_{\rm ph} = - (0.5 + 2i)$~meV, $\hbar\gamma_{\rm r} = 0.01$~meV, $ R = 0.01$~$\mu$m$\,$meV, $g_{\rm r}=0.225R$. An extended description of the model, along with the reasoning behind the parameter selection, can be found in Supplementary Section S3. 

\subsection*{The symmetry of the states}
The alternating symmetry of quantized states, manifested in measured  $k_{\parallel}$-space patterns, naturally follows from the intrinsic symmetry of the eigenvalue problem  \eqref{eq:eigenValueProblem}. 
For a symmetric pump profile $U(x) = U(-x)$, the Hamiltonian exhibits symmetry under the combined operation of spatial inversion $\hat{P}$ ($x\rightarrow -x$) and pseudo-spin flip $\hat{\sigma}_x$, which swaps rightward- and leftward-propagating polariton branches. This restricts the allowed single-particle polariton states to those satisfying $\vec{\Psi}(x) = \pm\sigma_x\vec{\Psi}(-x)$, which we associate with the symmetric (``$+$'') and antisymmetric (``$-$'') solutions.

The measured reciprocal state patterns reflect the distribution of the radiative losses and can be associated with the anti-Hermitian part $\hat{\mathcal{H}}_{\rm A} = -i\Gamma_{\rm p}+i\mathrm{Im}( V ) \hat{\sigma}_x$ of Hamiltonian \eqref{eq:hamilt}~\cite{Sigurdsson2024Dirac}. 
The $k_{\parallel}$-space emission intensity pattern is then governed by  $i \vec{\Psi}^\dagger(k_\parallel) \hat{\mathcal{H}}_{\rm A} \vec{\Psi}(k_\parallel)$. The reciprocal space patterns for the symmetric and antisymmetric solutions differ. In particular, at ${\rm Im}(V)>0$ the symmetric states have a minimum near $k_{\parallel}=0$, while the antisymmetric states have a peak. According to the fitting of the emission spectrum discussed in Supplementary Section S1, this is the case of the sample under study. The situation is reversed in the opposite case of negative dissipative coupling. 

\backmatter

\section*{Data Availability}
The data supporting the findings of this study are available within the Article and its Supplementary Information files as well as from the corresponding authors on reasonable request.

\section*{Acknowledgements}
The work was supported by the Russian Science Foundation Grant No. 25-72-20029.
A.N. and P.K. acknowledge support of “Basis” Foundation (Project No. 25-1-2-42-1).
S.M. acknowledges financial support from the National Natural Science Foundation of China (project 62350610272) and the Department of Science and Technology of Shandong Province (Grant KY0020240040). The authors thank Vladislav Petrenko for his contribution to the graphic design of the Figures.


\section*{Competing Interest Statement}
The authors declare no conflict of interest.

\noindent

\bibliography{sn-bibliography}

\end{document}